\documentclass[journal]{IEEEtran}

\IEEEoverridecommandlockouts
\usepackage{cite}
\usepackage{amsmath,amssymb,amsfonts}
\usepackage{algorithmic}
\usepackage{graphicx}
\usepackage{textcomp}
\usepackage{xcolor}
\def\BibTeX{{\rm B\kern-.05em{\sc i\kern-.025em b}\kern-.08em
    T\kern-.1667em\lower.7ex\hbox{E}\kern-.125emX}}

\graphicspath{{figures/}}
\usepackage[bookmarks,
    pdfauthor={Stephen McGovern},
    pdftitle={Spectral Processing of COVID-19 Time-Series Data},
    pdfsubject={COVID-19 data analysis},
    pdfkeywords={COVID-19, Coronavirus, SARS-CoV-2, smoothing, moving average, oscillation, spectral analysis, signal processing, synthesis, filtering, frequency domain},
    ]{hyperref}

\begin{document}

\title{Spectral Processing of \\ COVID-19 Time-Series Data\\
}

\author{\IEEEauthorblockN{Stephen McGovern}
\IEEEauthorblockA{
\textit{Wire Grind Audio}\\
Sunnyvale, United States \\
{\includegraphics[scale=0.35]{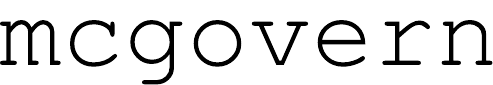}\hspace{0.2em}%
\includegraphics[scale=0.35]{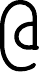}\hspace{0.2em}%
\includegraphics[scale=0.35]{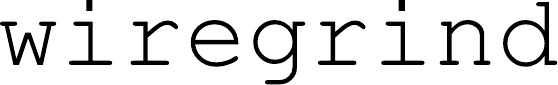}\hspace{0.2em}%
\includegraphics[scale=0.35]{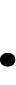}\hspace{0.2em}%
\includegraphics[scale=0.35]{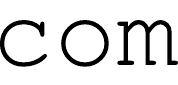}}\\
}
}

\maketitle

\begin{abstract}
The presence of oscillations in aggregated \mbox{COVID-19} data not only raises questions about the data's accuracy, it hinders understanding of the pandemic. Spectral analysis is used to reveal additional properties of the data, and the oscillations are replicated using sinusoidal resynthesis. The precise behavior of the seven-day moving average is also discussed, specifically, the cause of its jaggedness and the phase error it introduces. In comparison, other filtering techniques and Fourier processing produce superior smoothing and have zero phase error. Both of these are presented, and they are extended to isolate several frequency ranges. This extracts some of the same short-term variability that is resynthesized, and it shows that fluctuations with periods between 8 and 21 days are present in U.S. mortality data. These methods have applications that include modeling epidemiological time-series data as well as identifying less obvious properties of the data.
\end{abstract}

\vspace{2mm}


\section{Introduction}
\label{intro}

Multiple groups are aggregating data on COVID-19 including \cite{d1,d2,d3,d4}. By and large, they rely on a large number of other sources for their data. A comprehensive list of data sources is given by \cite{d1}. While the total number of daily cases and deaths being documented is without precedent, surveillance of other conditions has also produced comparable data. Daily data was produced during the Haitian cholera epidemic \cite{e0} as well as the West African Ebola epidemic \cite{e1}. Similarly, the World Health Organization and various governmental bodies track and published influenza data on a weekly basis.

For COVID-19, many countries and regions show clear oscillations in their daily counts of both new cases and new deaths. Over the course of just a few days, counts can experience dramatic fluctuations. For example, between July 6 and 7, the number of deaths in Arizona jumped from 1 to 117 \cite{d3}. On June 19th, Brazil had 55,209 new cases and then on June 21st, it had 16,851 \cite{d4}.

A number of studies have examined the oscillations \cite{e2,e3,b1,b2,b3,b4,b5,b7}. A couple of these point out that oscillations have been observed in prior epidemics \cite{e2,e3}. However, they also note that this occurred over a considerably longer time scale. It has been shown that the phases of the oscillations in different regions tend to align with one another more often than not \cite{b7}. Cases typically reach a maximum close to Friday before falling to a minimum close to Monday \cite{b2}.

A few different hypotheses have been proposed to explain the oscillations. New York City and Los Angeles county maintain their own data on COVID-19. According to \cite{b3}, in these datasets, dates are back-dated, and the oscillations are not present in their mortality data. It was also shown that oscillations in the number of infections resulted from oscillations in the number of tests being administered on a given day.

Even if the oscillations are caused by data acquisition practices, their presence is problematic for more than just mathematical analysis. They indicate that the data points contain a large amount of error. Identifying the sources of error could give insight into the shortcomings of data acquisition practices, and thereby lead to more effective pandemic response strategies.

There is an absence of any standard procedures for reporting and acquiring COVID-19 data.  Estimation of the total number of infections has been greatly hindered by tests not being available. A number of seroprevalence studies \cite{s1} have sought to help assess prevalence. Additionally, analyses of excess deaths have indicated that the true death toll is probably substantially higher than what is reported by data aggregators \cite{c1}.

The seven-day moving average has become an immensely popular method for suppressing oscillations in COVID-19 data. It's in widespread use by both news agencies and researchers. While superficially simple, its precise behavior is complex and unintuitive. Generally speaking, oscillations are attenuated rather than removed. It also flips the phase of oscillations if they lie within specific frequency ranges. Many papers have looked at trend analysis and forecasting. Some of these, such as \cite{a1,a2,a3}, have employed alternative smoothing methods such as exponential smoothing. In contrast, this paper employs  methodology from signal and audio processing theory. Spectral techniques are used for improved data smoothing as well as the extraction of shorter-term fluctuations. Additionally, spectral analysis and modeling techniques are used to resynthesize time-series oscillations. As a source of aggregated COVID-19 data, the repository from \cite{d1} is used.

\section{Spectral Analysis And Resynthesis}

Spectral analyses was performed in \cite{e3,b3,b4,b5} to study the 7-day oscillation.  Oscillations shorter than 7 days have also been observed by multiple researchers, such as \cite{e3,b3,b5}.

Our own spectral analysis found 7-day oscillations for both cases and deaths in many, but not all, countries. A number of countries exhibit a 3.5-day oscillation, and a few were found to have 2.33-day oscillations. Both of these time periods are integer divisors of 7 days; thus, they are potentially harmonics. Harmonics are frequently observed in physical systems, and they are always present in more complex periodic waveforms (e.g., square waves and triangle waves).

\begin{figure*}[t]
\centerline{\includegraphics[scale=0.70]{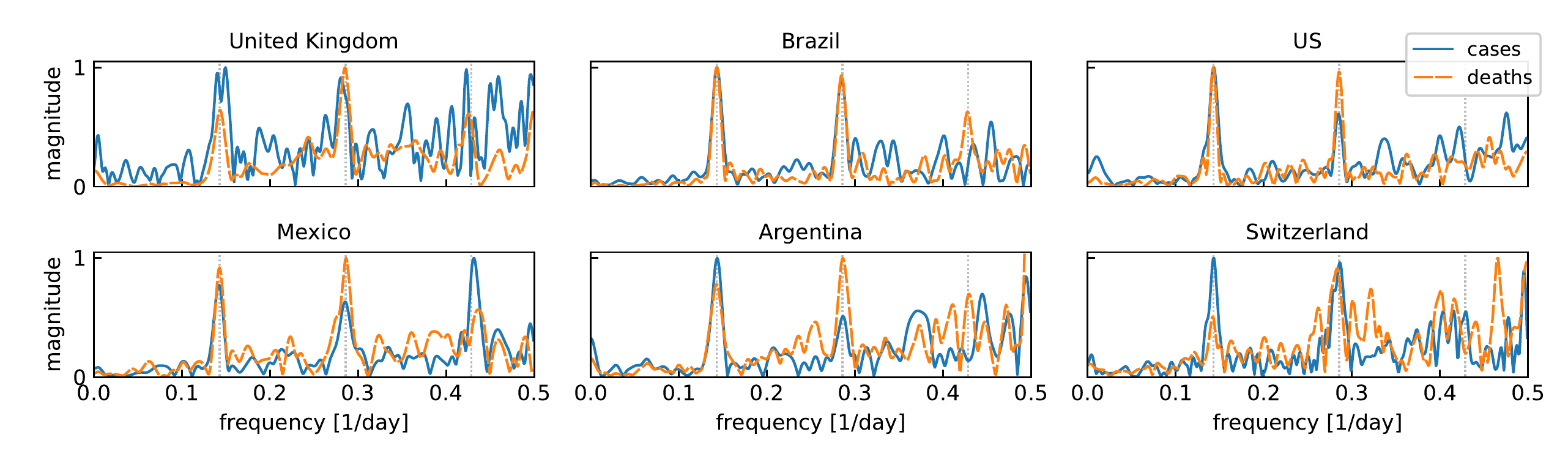}}
\vspace*{-3mm}
\caption{\label{fig:spectrum}{\it Spectrograms of the derivatives of daily counts for both cases and deaths. Vertical dotted lines mark periods of 7.0, 3.5, and 2.33 days from left to right. Calculations use the most recent 193 days of data and a Hanning window function. Magnitudes are normalized to the largest value between the frequencies 0.1/day and 0.475/day.}}
\vspace*{3mm}
\end{figure*}

Three papers \cite{e3,b1,b4} used first differences in their spectral analyses. Conceptually, a first difference is very similar to a derivative. Both amplify higher frequencies and attenuate lower ones. First differences have a phase response that varies linearly with frequency, while true derivatives have a phase response that is constant. For time-series data, attenuation of low frequencies has the effect of removing long-term trends. For COVID-19 data, it also makes the peaks for the 3.5 and 2.33-day oscillations more prominent relative to the peak for 7-day oscillations. Spectrograms of the derivatives are shown in Fig. \ref{fig:spectrum} for six countries.

\begin{figure*}[t]
\centerline{\includegraphics[scale=0.70]{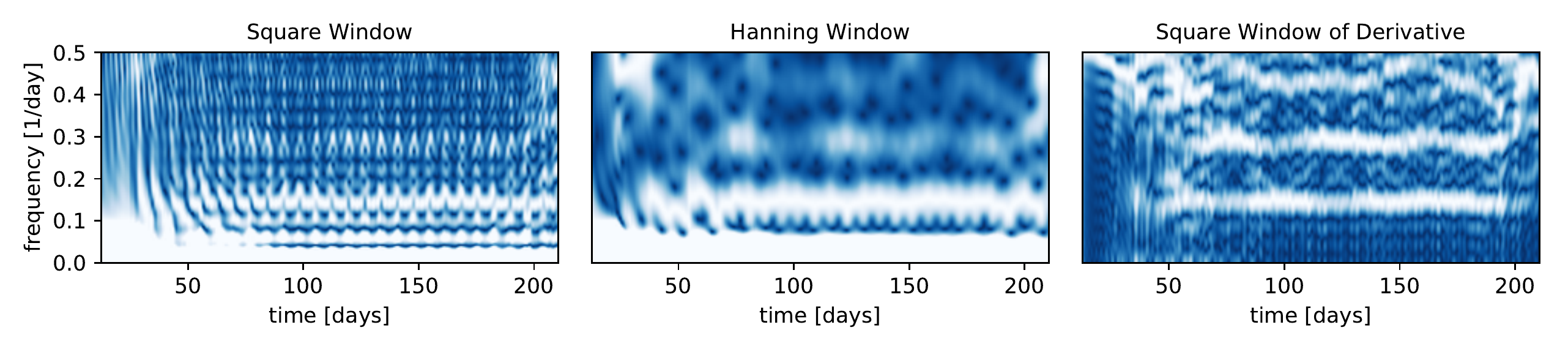}}
\vspace*{-3mm}
\caption{\label{fig:time_spectrum}{\it Time-dependent spectra of Brazil's new daily deaths using a 25-day sliding window. For each time-window, the spectra are normalized to the largest value above the frequency of 0.1/day. Below 0.1/day, values exceeding unity are clipped.}}
\vspace*{3mm}
\end{figure*}

As of this writing, there are only around 185 days in the COVID-19 time-series that are mathematically significant to this paper's analysis. From a signal processing perspective, this is an extraordinarily small number of data points. Sections of the daily counts also contain exponential growth and decay. Selecting different time-ranges can cause the locations of spectral peaks to wobble and sometimes disappear altogether. Applying window functions is also problematic as data becomes blurred, leaving spectral features more difficult to discern. It was observed that when Fourier transforms were instead applied to the derivatives of the time-series, spectral variations were reduced and features became more clear. This is illustrated in Fig. \ref{fig:time_spectrum} using a sliding time-window.

In this paper, derivatives are calculated in the frequency domain. Time-domain calculations can also work; however, various spectral, phase, and time-shifting artifacts generally occur. Depending on the particular use case, such artifacts might need to be accounted for. A brief comparison of several differentiation methods is found in Appendix \ref{sec:CompareDeriv}. Methodology for computing frequency domain derivatives is outlined in Appendix \ref{sec:FreqDomainDeriv}.

\subsection{Resynthesis}
\label{sec:resynth}
Reproducing time-series oscillations can aid in understanding the weekly progression of the epidemic. It can be used to forecast minimums, maximums, and plateaus in the daily counts of infections, deaths, tests, and other data. It can also help identify varying data acquisition practices, and it's potentially useful for detecting irregularities in data.

Given a time-series $x[n]$ and its $N$-point Fourier transform $X[k]$, the magnitude and phase angle of the frequency components are respectively given by,

\begin{equation}
a[k] = \frac{2}{N} \Big| X[k] \Big|,
\label{eq:magnitude}
\end{equation}

\begin{equation}
\theta [k] = \tan^{-1} \left( \frac{ \operatorname{Im} (X[k]) }{ \operatorname{Re} (X[k]) } \right).
\label{eq:angle}
\end{equation}

\noindent Using Eqs. \ref{eq:magnitude} and \ref{eq:angle}, and zero-indexed arrays, the time-series can then be reconstructed as,

\begin{equation}
x[n] = \sum_{k=0}^{N/2-1} a[k] \cos{ \left(  \frac{2\pi nk}{N} + \theta[k] \right)}.
\label{eq:reconstruct}
\end{equation}

Using sinusoidal resynthesis, the oscillations were recreated for the three aforementioned frequencies. The derivative of the time-series was taken, followed by the Fourier transform. Then Eq. \ref{eq:reconstruct} was symbolically integrated with respect to $n$, and the summation was taken only for values of $k$ that corresponded to one of the three frequencies. The result is a waveform having only the oscillations in the original time-series.

\begin{figure*}[t]
\centerline{\includegraphics[scale=0.70]{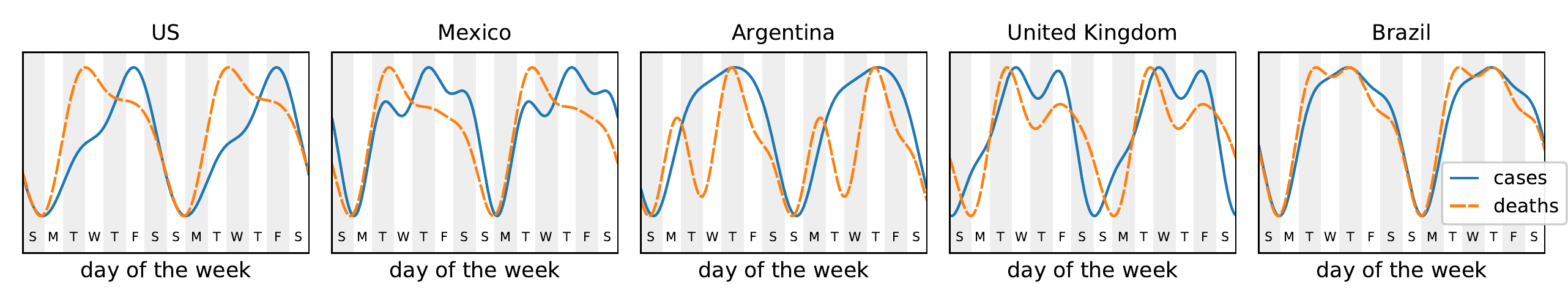}}
\vspace*{-3mm}
\caption{\label{fig:waveforms}{\it Resynthesis of the time-series' three oscillations using a 183-day analysis window.}}
\vspace*{3mm}
\end{figure*}

The input dataset is aggregated on a daily basis. To improve time alignment, each data point was treated as having occurred at 12 noon on its respective day. Resynthesis calculations were then performed using minute-level time-resolution, and the result of this is shown in Fig. \ref{fig:waveforms} for five countries. The waveform shapes can change overtime, and revisions are often made to existing data points.

\section{Smoothing and Isolation}
The behavior of COVID-19 time-series data can be better understood and better characterized by altering its spectral properties. In fact, spectral alteration is the precise mechanism by which the seven-day moving average smooths data. In this study, three methods were used for modifying the spectral content of U.S. time-series data: the moving average filtering, infinite impulse response (IIR) filtering, and frequency domain processing. As an important first and final step, the data underwent pre- and post-processing.

\subsection{Properties of the Seven-Day Moving Average}
\label{sec:movingAverage}

In the context of digital signal processing, the seven-day moving average is a finite impulse response (FIR) filter. Its time-center form is non-causal, and it can be denoted by,

\begin{equation}
y[n] = \frac{1}{7} \sum_{k=-3}^{3} x[n-k],
\label{eq:MovAve}
\end{equation}

\noindent where $x$, $y$, and $n$ are the input, the smoothed output, and the time-index, respectively. Taking the Z-transform, rearranging, and simplifying yields the filter's transfer function,

\begin{equation}
H(z) = \frac{1}{7} \left( z^{-3} + z^{-2} + z^{-1} + z^{0} + z^{1} + z^{2} + z^{3} \right).
\label{eq:MovAveTf}
\end{equation}

\noindent Substituting $e^{j 2 \pi f}$ for $z$ in Eq. \eqref{eq:MovAveTf} and taking the absolute value results in the frequency response. It follows that the frequency response in decibels is given by,

\begin{equation}
H_{dB}(f) = 20 \log_{10}\left|{H}\left(e^{j 2 \pi f}\right)\right|,
\label{eq:MovAveTfDb}
\end{equation}

\noindent where $f$ is the frequency. The continuous frequency phase response is obtained by substituting ${H}\left(e^{j 2 \pi f}\right)$ for $X[k]$  in Eq. \ref{eq:angle}.

\begin{figure}[t]
\centerline{\includegraphics[scale=0.70]{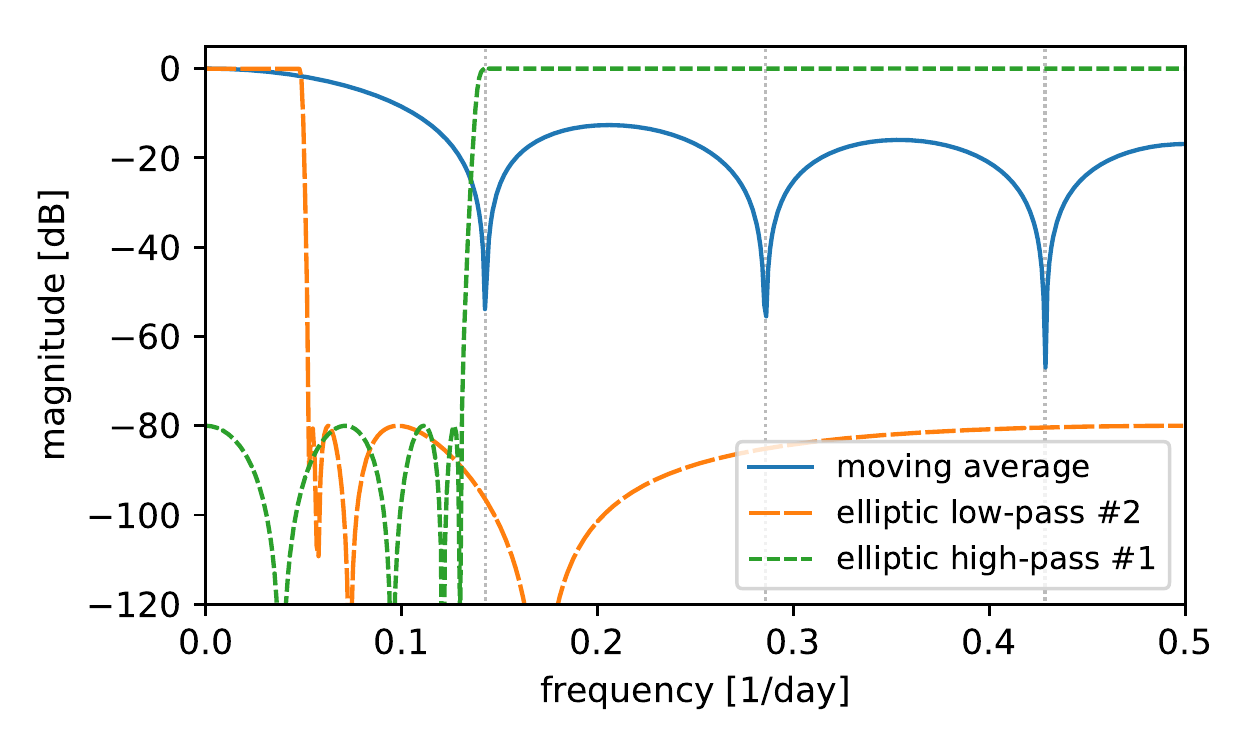}}
\vspace*{-3mm}
\caption{\label{fig:FiltComp}{\it The frequency response for three filters. Vertical dotted lines mark periods of 7.0, 3.5, and 2.33 days from left to right.}}
\vspace*{3mm}
\end{figure}

\begin{figure}[t]
\centerline{\includegraphics[scale=0.70]{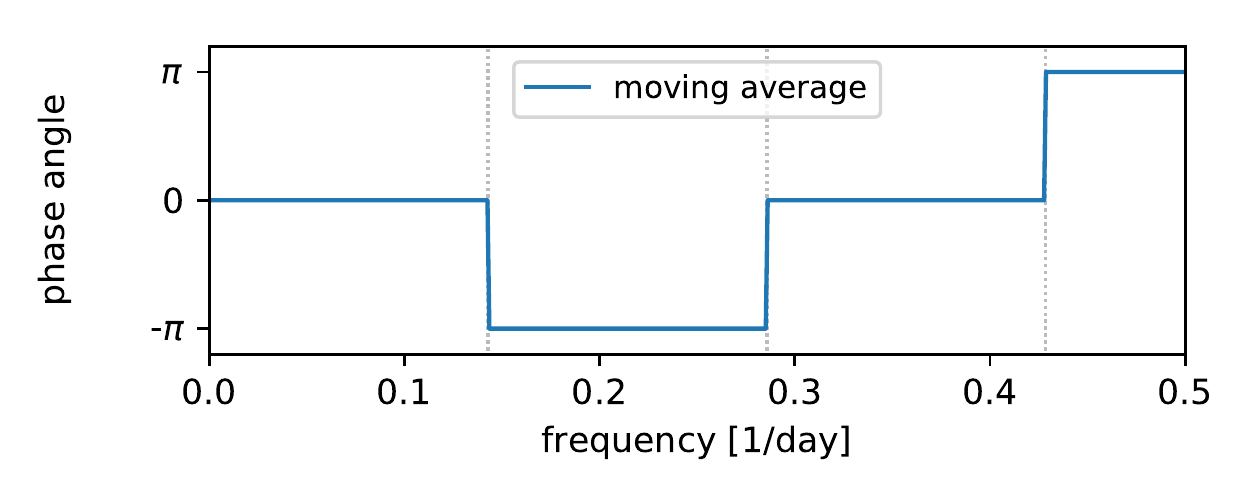}}
\vspace*{-3mm}
\caption{\label{fig:MaPhase}{\it The phase response of a seven-day moving average. Vertical dotted lines mark periods of 7.0, 3.5, and 2.33 days from left to right.}}
\vspace*{3mm}
\end{figure}

\begin{figure}[t]
\centerline{\includegraphics[scale=0.70]{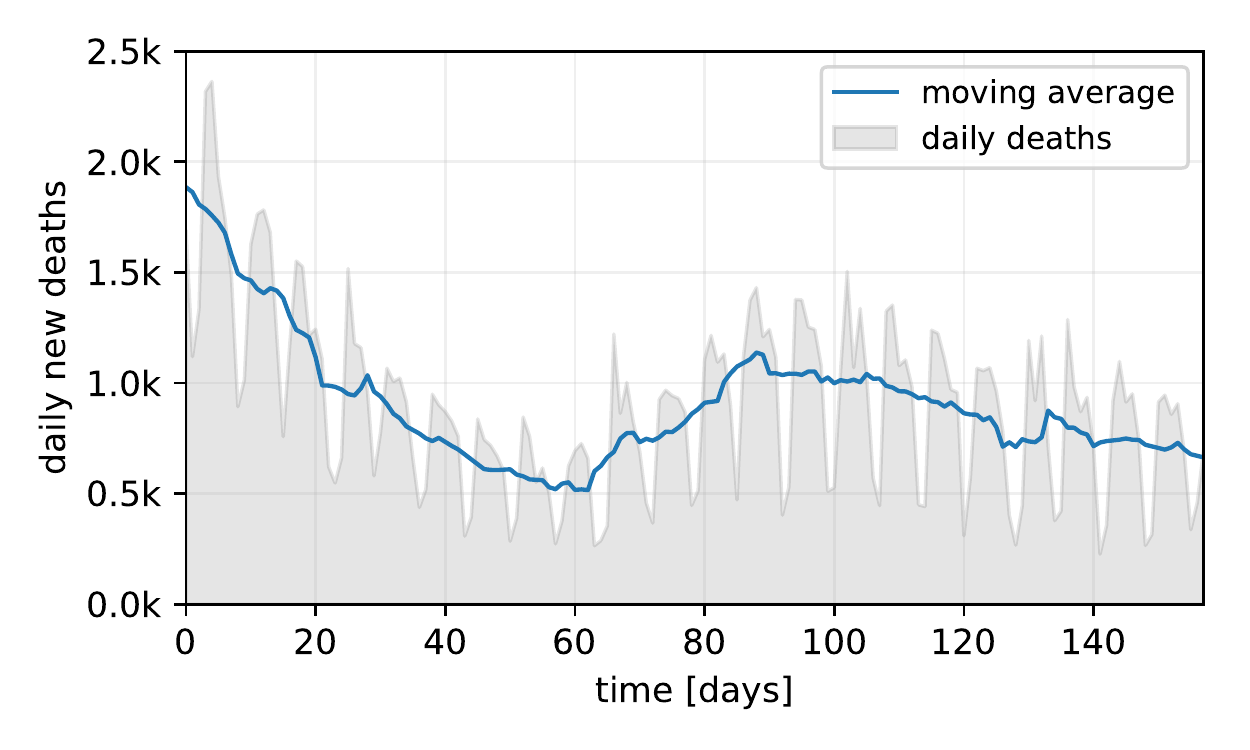}}
\vspace*{-3mm}
\caption{\label{fig:MovingAverage}{\it New daily deaths in the U.S. before and after processing. The seven-day moving average leaves the plot somewhat jagged.}}
\vspace*{3mm}
\end{figure}

The frequency response for the seven-day moving average is shown in Fig. \ref{fig:FiltComp}. The three nulls in the plot occur for frequencies with periods of 7 days, 3.5 days, and 2.33 days. The presence of nulls indicates that the filter will completely remove the respective frequencies. All other frequencies, including those periods less than 7 days, are still present. While they are reduced in magnitude, it is enough to be the sole cause of the jaggedness seen in media reports. The jaggedness is illustrated in Fig. \ref{fig:MovingAverage}. To further muddle the data, frequencies on certain intervals have inverted phases. This is illustrated in the phase response in Fig. \ref{fig:MaPhase}. A second application of the same moving average filter will correct the phase inversions.

\subsection{Processing Methodology}

\subsubsection{Pre- and Post-Processing}
Padding the input dataset is beneficial to all three spectral processing methods. It allows the moving average to run to the end of the time-series. It facilitates the initialization of the IIR filters, and it can help prevent artifacts that frequency domain processing would otherwise leave at the ends of the modified time-series.

Before processing, 28 days of synthetic data were added to both ends of the original data. Each synthetic data point was calculated by linearly extrapolating from the nearest existing data points located at distances of $7m$ and $7(m+1)$ days where $m$ is an integer. Numbers extrapolating to less than $0$ were set equal to $0$. After processing, sections of the output data corresponding to extrapolated data were removed.

\subsubsection{IIR Filters}
Elliptic filters are a type of IIR filter. Their steep frequency roll-off permits strong frequency isolation. Frequency-dependent phase changes are a common artifact of signal processing filters. However, these artifacts are canceled out if filtering is applied twice in opposite directions.

Four elliptic filters were created using Python's SciPy library. Comparable functionality is found in Matlab and Octave. The dataset was filtered twice: backwards first and then forwards.  The filter design parameters are given in Table \ref{tab:params}. Frequency response plots for two passes of two of the listed filters are shown in Fig. \ref{fig:FiltComp}.

\begin{table*}[t]
\caption{\label{tab:params} Parameters For Elliptic Filters}
\begin{center}
\begin{tabular}{|c|c|c|c|c|}
\hline
filter name & pass-band frequency & stop-band frequency & pass-band ripple$^{\mathrm{a}}$ & stop-band attenuation$^{\mathrm{a}}$ \\
\hline
\textbf{low-pass \#1} & 1/9      & 1/8     & 0.01 dB & 40 dB  \\
\textbf{low-pass \#2} & 1/21     & 1/19    & 0.01 dB & 40 dB  \\
\textbf{high-pass \#1}    & 1/7      & 1/8     & 0.01 dB & 40 dB  \\

\textbf{band-pass \#1}    &
\begin{tabular}{cc}1/8 & 1/6\end{tabular} &
\begin{tabular}{cc}1/9 & 1/5\end{tabular} &
0.01 dB & 40 dB \\

\textbf{band-pass \#2}    &
\begin{tabular}{cc}1/19 & 1/9\end{tabular} &
\begin{tabular}{cc}1/21 & 1/8\end{tabular} &
0.01 dB & 40 dB  \\

\hline
\multicolumn{5}{l}{$^{\mathrm{a}}$Two filter passes result in total values of $0.02$ dB and $80$ dB.}
\end{tabular}
\end{center}
\end{table*}

\subsubsection{Frequency Domain Processing}
The Fourier transform will convert a time-series to and from the frequency domain. From there, its spectral content is readily modified. The general formula for such operations can be written as,

\begin{equation}
y[n] = \mathcal{F}^{-1}( H_{s}[k] \cdot \mathcal{F}( x[n] ) ),
\label{eq:FFT}
\end{equation}

\noindent where $\mathcal{F}$, $\mathcal{F}^{-1}$, and $H_{s}[k]$ are the Fourier transform, the inverse Fourier transform, and a computed spectrum, respectively.

Five spectra were computed for $H_{s}[k]$. A "brick wall" spectrum could be calculated by setting $H_{s}[k]$ equal to one or more unit step sequences mirrored about Nyquist. For low-pass, high-pass, and band-pass spectra, this can implemented using the formula,

\begin{equation}
    H_{s}[k] =
\begin{cases}
    1,  & \text{if } N(f_{low}) \leq k \leq N(f_{high})  \\
    1,  & \text{if } N(1-f_{high}) \leq k \leq N(1-f_{low})  \\
    0,  & \text{otherwise},
\end{cases}
\label{eq:brick_wall}
\end{equation}

\noindent  where $f_{low}$ and $f_{high}$ are the lower and upper bounds of the pass-band. However, for reasons that will become more clear in Sec. \ref{results}, "brick wall" spectra were closely approximated by,

\begin{equation}
H_{s}[k] = \left| {H_{e}[k]}^2 \right|,
\label{eq:spectra}
\end{equation}

\noindent where $H_{e}[k]$ is the discretely sampled transfer function of the single-pass filters described in Table \ref{tab:params}.

\subsection{Results and Discussion}
\label{results}

The moving average was applied to the COVID-19 data. The result is plotted in Fig. \ref{fig:MovingAverage}. Using the parameters from Table \ref{tab:params}, the elliptic filters and the frequency domain method were also used to process the data.

\begin{figure}[t]
\centerline{\includegraphics[scale=0.70]{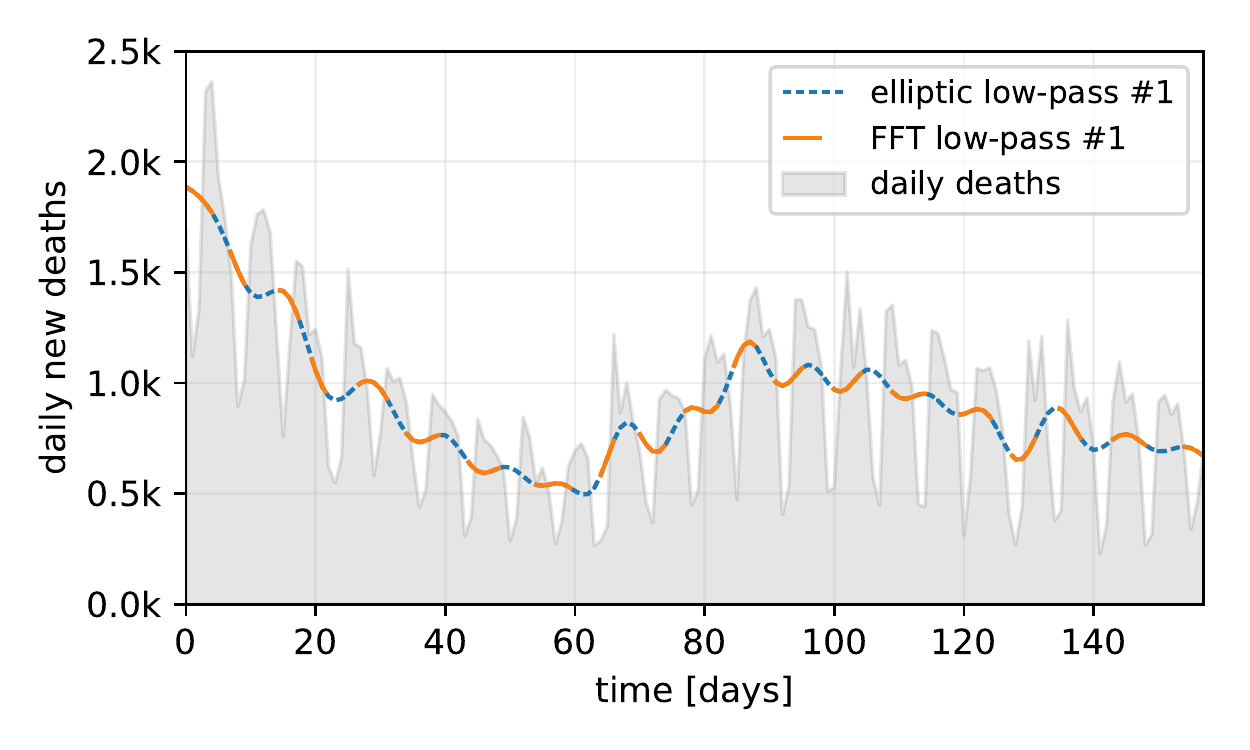}}
\vspace*{-3mm}
\caption{\label{fig:LP1}{\it New daily deaths in the U.S. before and after processing. Low-passing removes all oscillations shorter than 8 days. However, longer-term fluctuations are still visible.}}
\vspace*{3mm}
\end{figure}

\paragraph{\textbf{Low-pass \#1} (Fig. \ref{fig:LP1})}
\label{sec:LP1}
The line is visibly less jagged than the moving average. However, significant oscillations longer than $8$ days are still present. The elliptic filters and the Fourier method are in almost perfect agreement with one another.

\begin{figure}[t]
\centerline{\includegraphics[scale=0.70]{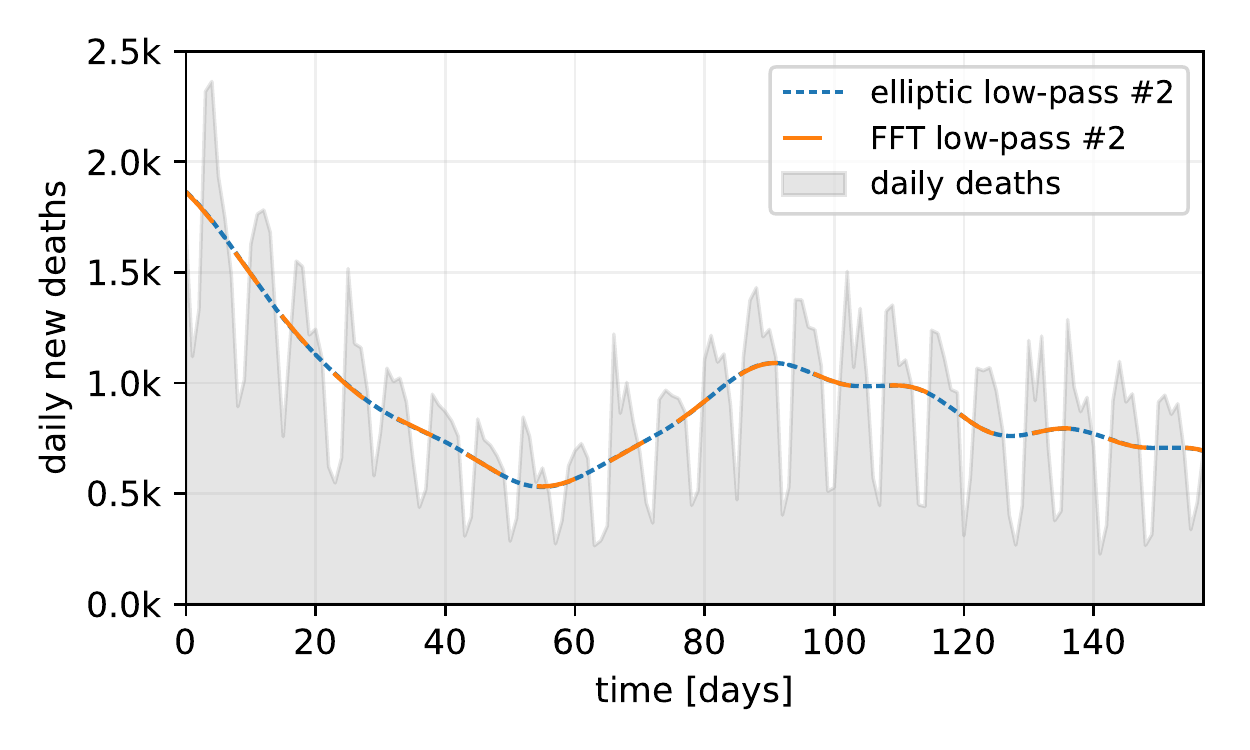}}
\vspace*{-3mm}
\caption{\label{fig:LP2}{\it New daily deaths in the U.S. before and after processing. Low-passing removes short-term oscillations.}}
\vspace*{3mm}
\end{figure}

\paragraph{\textbf{Low-pass \#2} (Fig. \ref{fig:LP2})}
The line is again less jagged. Also, there are fewer lower-frequency oscillations. As in Fig. \ref{fig:LP1}, the elliptic filters and the Fourier method are in almost perfect agreement.

\begin{figure}[t]
\centerline{\includegraphics[scale=0.70]{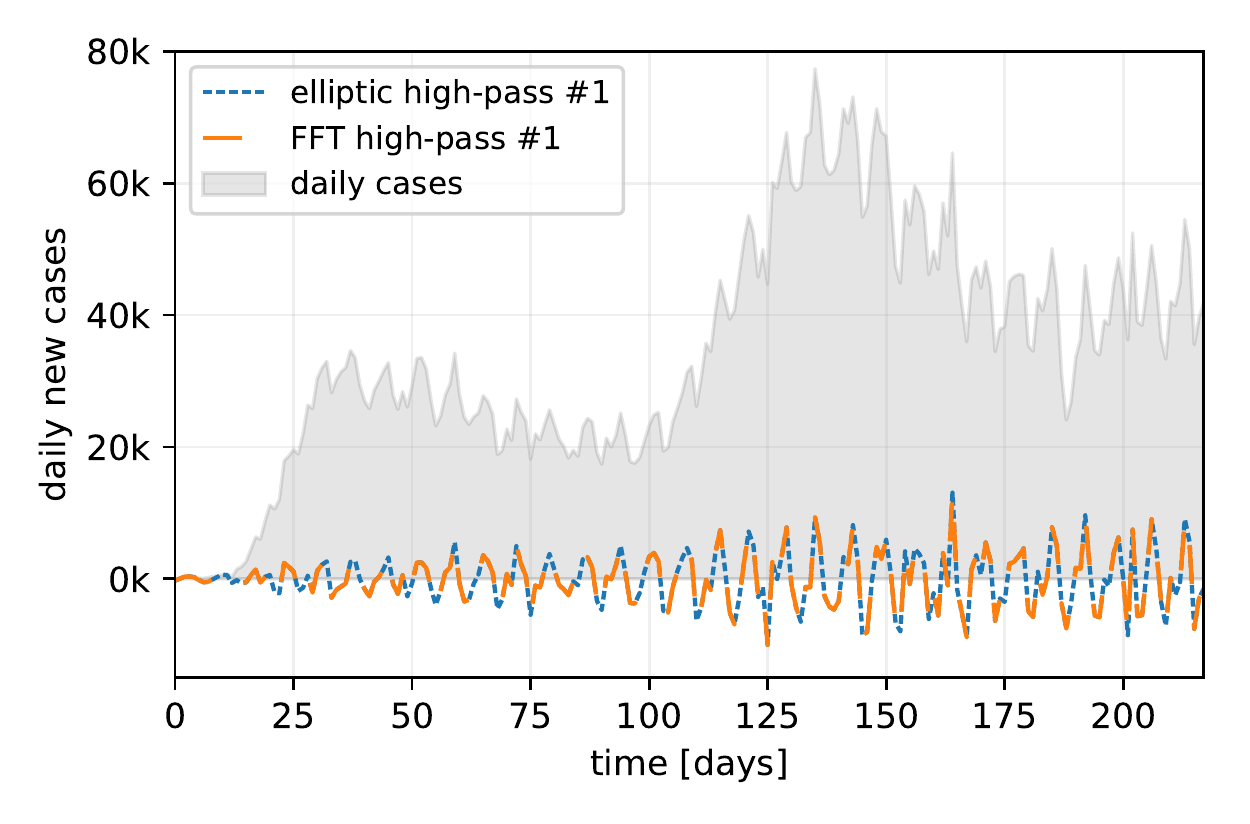}}
\vspace*{-3mm}
\caption{\label{fig:HP1}{\it New daily cases in the U.S. before and after processing. High-passing removes long term trends.}}
\vspace*{3mm}
\end{figure}

\paragraph{\textbf{High-pass  \#1} (Fig. \ref{fig:HP1})}
The long-term trends are effectively removed. As in the two low-pass methods, the elliptic filters and the Fourier method are once again in near perfect agreement.

\begin{figure}[t]
\centerline{\includegraphics[scale=0.70]{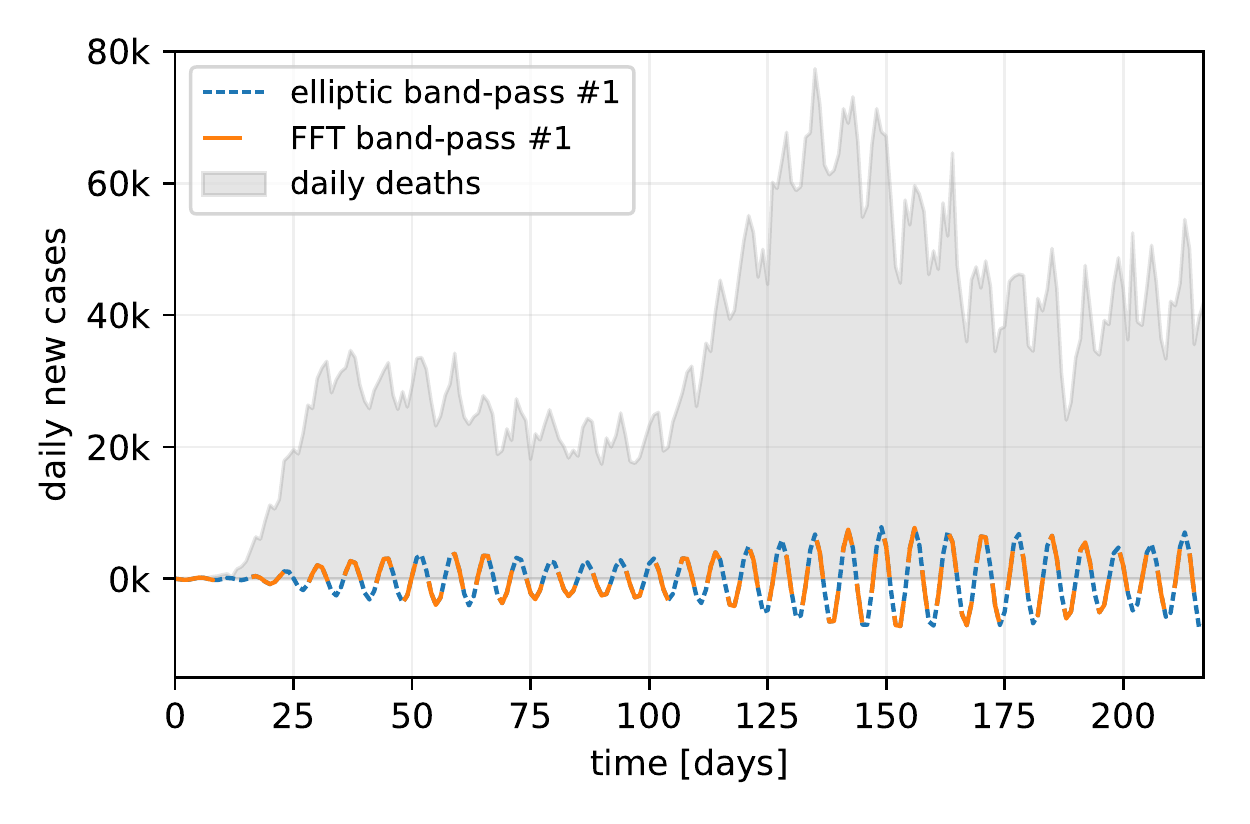}}
\vspace*{-3mm}
\caption{\label{fig:BP1}{\it New daily cases in the U.S. before and after processing. Band-passing isolates the oscillations between 6 and 8 days.}}
\vspace*{3mm}
\end{figure}

\paragraph{\textbf{Band-pass \#1} (Fig. \ref{fig:BP1})}
The 7-day oscillation is fairly well isolated. The resultant waveform is much more sinusoidal in nature than that produced by the high-pass filter. Like the other operations, the two methods agree very well.

\begin{figure}[t]
\centerline{\includegraphics[scale=0.70]{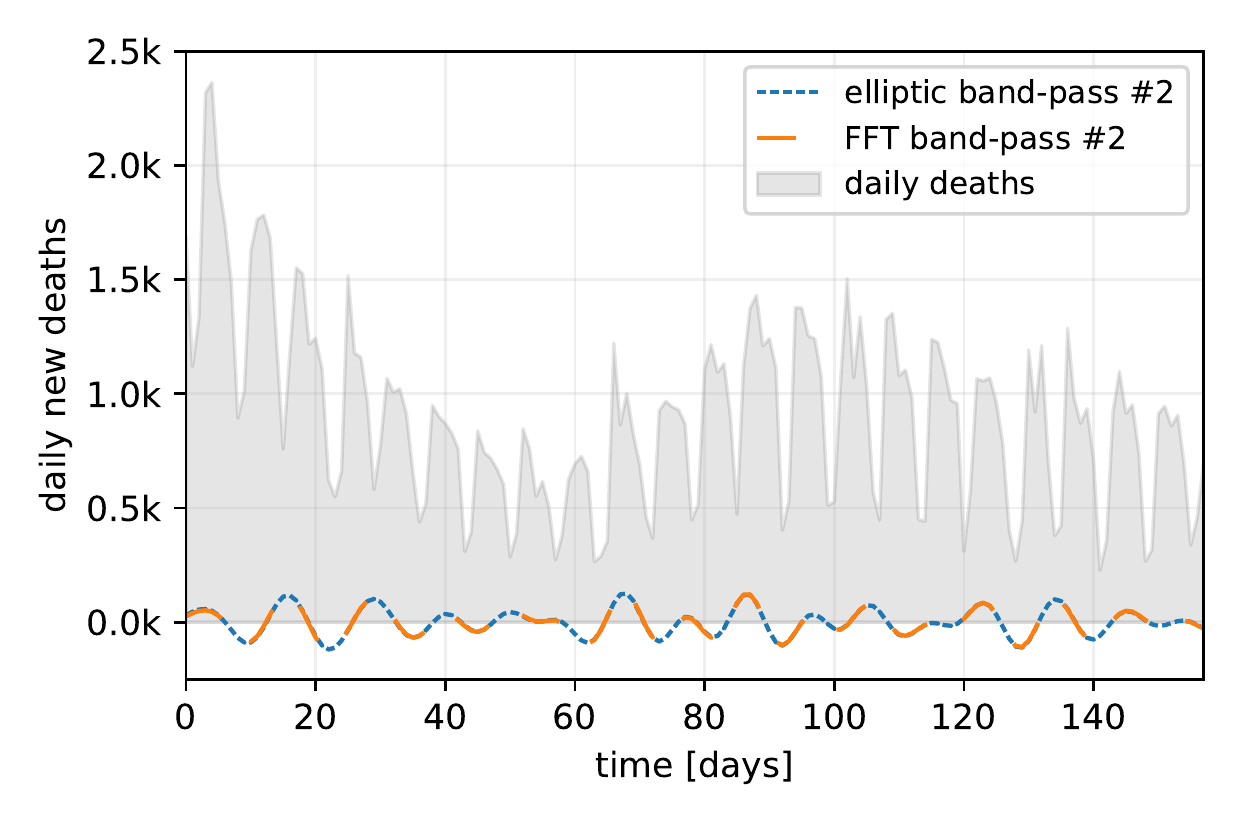}}
\vspace*{-3mm}
\caption{\label{fig:BP2}{\it New daily deaths in the U.S. before and after processing. Band-passing isolates oscillations with periods between 8 and 21 days.}}
\vspace*{3mm}
\end{figure}

\paragraph{\textbf{Band-pass \#2} (Fig. \ref{fig:BP2})}
A bandwidth of oscillations with periods between 8 and 21 days is separated. Aperiodic oscillations remain despite having 80 dB of attenuation on all oscillations with periods shorter than 8 days. The remaining oscillations are the same as those that were present in Fig. \ref{fig:LP1} but not present in Fig. \ref{fig:LP2}. The two methods still produce plot lines that overlay one another, albeit with slightly less precision than in the other trials.

\noindent \newline One thing to note here, is that the temporal resolution worsens as the band-width is narrowed. A principle of Fourier theory is that an arbitrary time-series can be represented completely by a summation of sinusoids. Thus, if some of the sinusoids are removed from the series, there could be oscillations in locations where the input series was near zero. Those oscillations are in fact present in the input; they are just not necessarily visible when the other frequency components are also present. A practical example of this is resonance in the earlier part of the data where there is a small number of cases, for example, the period prior to day 10 on Figs. \ref{fig:HP1} and \ref{fig:BP1}. On the other hand, the time period prior to day 10 is of the least interest.

\section{Conclusions}

Significant spectral differences were observed among different countries. These differences are often evident in the time-series data. For example, several countries had a large peak for 3.5-day oscillations in at least one of their spectra. Upon reviewing the respective time-series, a double-humped oscillation was observed. Provided that there is no biological explanation, this must result from some kind of tangible difference in how information is being collected, reported, and aggregated.

The oscillations in the time-series were recreated using sinusoidal resynthesis. This effectively recreated the minimums, the maximums, and the general shape of the time-series data including double-humps. This gives us some ability to predict the behavior of the time-series over the course of a week.

IIR filters and frequency domain processing can be used to selectively manipulate the spectral properties of COVID-19 time-series data. When tuned appropriately, both methods produce virtually identical results. The superior suppression of higher-frequency components smooths data far more effectively than the seven-day moving average. It would be difficult to argue that statistically sensitive calculations should carried out using a statistically erroneous seven-day moving average. Additionally, the isolation of higher-frequency components could be useful for predicting and modeling short-term variations in observed data. Using the described methods, fluctuations greater that 8 days and shorter than 21 days were identified in U.S. mortality data. The cause and significance of this is currently not known.


\section*{Appendix A: Numerical Derivatives}
\label{sec:CompareDeriv}

Many methods exist for approximating derivatives from discretely sampled data. These include first differences, central differences, and frequency domain derivatives. The first difference is given by

\begin{equation}
x'[n] = x[n] - x[n-1],
\label{eq:firstDiff}
\end{equation}

\noindent This is a backwards method, and it produces a time-shifting artifact of $\frac{1}{2}$ sample period. More accurate estimates can be achieved with higher-order methods. A central difference formula is given in Eq. \ref{eq:centralDiff}.

\begin{equation}
\begin{split}
x'[n] = &   (1/280)x[n-4] - (4/105)x[n-3] \\
        & +   (1/5)x[n-2] -   (4/5)x[n-1] \\
        & +   (4/5)x[n+1] -   (1/5)x[n+2] \\
        & + (4/105)x[n+3] - (1/280)x[n+4]
\label{eq:centralDiff}
\end{split}
\end{equation}

\begin{figure}[t]
\centerline{\includegraphics[scale=0.70]{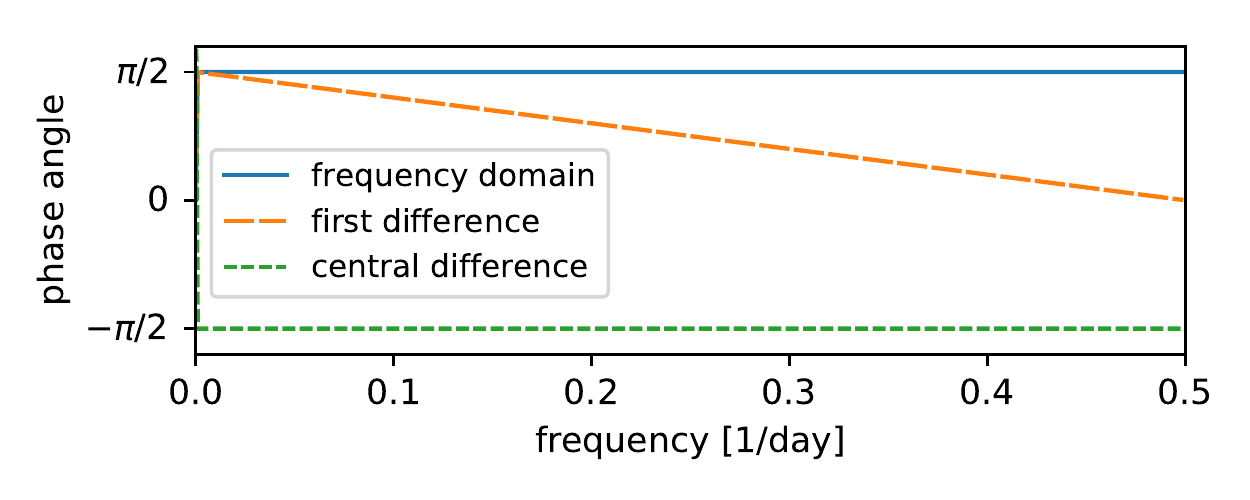}}
\vspace*{-3mm}
\caption{\label{fig:detrendPhase}{\it Phase response of three numerical differentiation methods.}}
\vspace*{3mm}
\end{figure}

\begin{figure}[t]
\centerline{\includegraphics[scale=0.70]{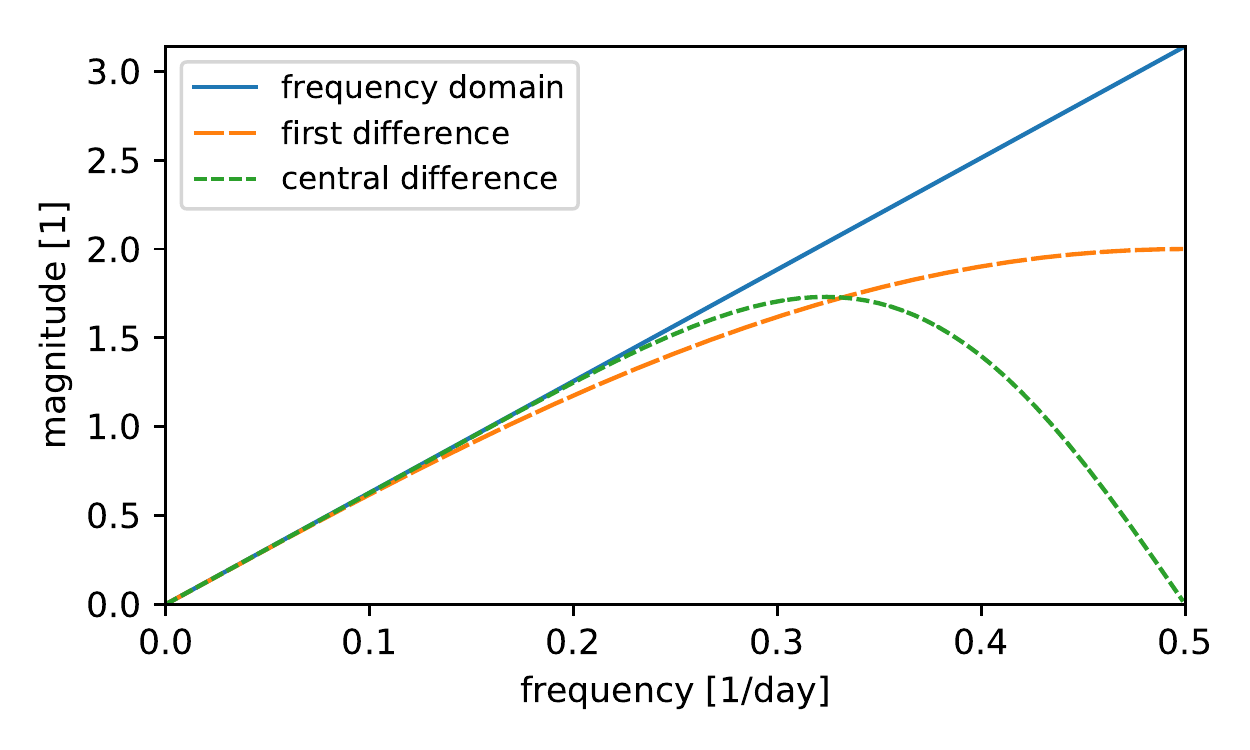}}
\vspace*{-3mm}
\caption{\label{fig:detrendSpec}{\it Spectral response of three numerical differentiation methods.}}
\vspace*{3mm}
\end{figure}

\noindent Due to the symmetry about $n$, Eq. \ref{eq:centralDiff} produces no time-shifting artifact. It is also a higher-order formula. Frequency domain derivatives exist as yet another method.  See Appendix \ref{sec:FreqDomainDeriv} for a mathematical description. For these three methods, the phase and spectral responses are plotted in Figs. \ref{fig:detrendPhase} and \ref{fig:detrendSpec}, respectively.

\section*{Appendix B: Frequency Domain Derivatives}
\label{sec:FreqDomainDeriv}
The $N$-point spectrum of the complex differentiation operation can be given as,

\begin{equation}
    H_{\frac{d}{dn}}[k] =
\begin{cases}
    j 2 \pi (k/N),     & \text{if } 0 \leq k < N/2 \\
    j 2 \pi (k/N - 1),  & \text{if } N/2 \leq k < N .
\end{cases}
\label{eq:fft_deriv0}
\end{equation}

\noindent Multiplying $H_{\frac{d}{dn}}[k]$ by the Fourier transform of the input results in the frequency domain representation of the derivative, $X'[k]$. This is given as,

\begin{equation}
X'[k] = H_{\frac{d}{dn}}[k] \hspace{1mm} \mathcal{F} ( x[n] ),
\label{eq:fft_deriv1}
\end{equation}

\noindent where $\mathcal{F}$ is the Fourier transform and $x[n]$ is the time-series. It follows that the derivative of $x[n]$ is then given by

\begin{equation}
x'[n] = \mathcal{F}^{-1}( X'[k]),
\label{eq:fft_deriv2}
\end{equation}

\noindent where  $\mathcal{F}^{-1}$ is the inverse Fourier transform.

\end{document}